\documentclass[preprint,pra]{revtex4}
\usepackage{hyperref}
\usepackage{amsmath}
\usepackage{amssymb}
\usepackage[pdftex]{graphicx}
\usepackage{microtype}

\newcommand{\mrm}{\mathrm}

\newcommand{\Esca}{\mathcal{E}}

\newcommand{\Bsca}{\mathcal{B}}
\newcommand{\sd}{$^1S_0 \to {}^1D_2$ \ }

\begin{document}
\title{Optical frequency standard based on a two-photon transition in calcium}
\author{Amar C. Vutha}
\affiliation{Department of Physics and Astronomy, York University, 4700 Keele St, Toronto ON M3J 1P3, Canada}
\email{avutha@yorku.ca}

\begin{abstract}
A two-photon transition in laser-cooled and trapped calcium atoms is proposed as the atomic reference in an optical frequency standard. An efficient scheme for interrogation of the frequency standard is described, and the sensitivity of the clock transition to systematic effects is estimated. Frequency standards based on this transition could lead to compact and portable devices that are capable of rapidly averaging down to $< 10^{-16}$.
\end{abstract}

\maketitle

\section{Introduction}

With their extremely high quality factors, narrow optical resonances in atoms are ideal candidates for realizing a highly stable atomic frequency reference. Optical frequency standards (OFS), consisting of narrow linewidth lasers stabilized to atomic transitions, have improved their performance significantly over the last decade. They are soon likely to lead to a more accurate re-definition of the SI second \cite{Gill2005,Poli2014}. The atoms in these frequency standards must be isolated from external perturbations and frequency shifts due to atomic motion, and therefore OFS typically use trapped atoms or atomic ions. The best performance to date has been obtained by interrogating the narrow $^1S_0 \to {}^3P_0$ transition in Sr and Yb atoms trapped in optical lattices, to eliminate Doppler and photon recoil shifts, with the lattice lasers tuned to a ``magic'' wavelength to minimize perturbations on the atoms \cite{Ludlow2014,Margolis2014}.

Aside from the primary optical clocks that will form the basis of the new SI second, there is a need for ensembles of secondary frequency standards to provide optical flywheels for generating timescales \cite{Parker2012}. A compact and transportable standard could also address the important problem of comparing the performance of widely separated primary frequency standards. In addition to improving the accuracy of atomic timekeeping, secondary OFS will also find applications in low-noise microwave synthesis (using frequency combs to transfer the phase stability of optical waves to microwaves \cite{McFerran2005}) and precise geodetic surveys (using the high sensitivity of optical clocks to gravitational red-shifts \cite{Chou2010}). An array of high-performance frequency standards aboard satellites could also be sensitive to gravitational waves \cite{Smarr1983,Armstrong2006}. However, all these applications need a compact OFS that can provide robust long-term performance. A simple design with low system complexity will be an important step towards this goal.

In this paper, the $E1^2$ two-photon transition between the $4s^2 \ {}^1S_0 \to 4s 3d \ {}^1D_2$ states in calcium atoms is offered as a means to realize a compact optical frequency standard. The clock transition between the $m=0$ sublevels of the $^1S_0$ and $^{1}D_2$ states is insensitive to magnetic fields. The two-photon clock transition, driven with identical counter-propagating photons, is free from 1st-order Doppler shifts and photon recoil. This transition, and its analogs in the other alkaline earth atoms, were examined in \cite{Hall1989} as a possible means of realizing optical fountain clocks. However, the realization of fountain clocks is complicated by large 2nd-order Doppler shifts in the beam of atoms, and the large angular divergence of the atomic beams after single-stage laser-cooling. In this work, we show that magneto-optical trapping of calcium atoms circumvents the woes associated with fountain clocks, and leads to a simple scheme to realize an optical frequency standard. Among the alkaline earth atoms, the feasibility of this scheme is special to Ca, as the ${}^1D_2$ state in Mg is not metastable, whereas the lifetime of the $^1D_2$ state in Sr and Ba are significantly shorter than in Ca. Compared to optical lattice clocks, this simplified scheme eliminates the second-stage cooling lasers and MOTs, the lattice laser, and the associated vacuum and optical hardware from the apparatus. This scheme is therefore well-suited to the realization of portable and low-maintenance secondary optical standards capable of high stability and accuracy. In the following section, the design of such a frequency standard, and an efficient interrogation scheme for the clock transition, are described. This is followed by estimates of the susceptibility of the clock transition to undesired frequency shifts. 

We note in passing that an even narrower two-photon transition is generally available in the alkaline earth atoms between the $^1S_0 \to {}^3D_2$ states. However, using this as the basis for an optical frequency standard requires a higher-power clock laser (which leads in turn to larger light shifts), as well as more lasers to repump out of the $^3D$ states. In addition, the proximity of the $^3D_1$ and $^3D_{0,2}$ states leads to 2nd order Zeeman shifts that are larger than for the \sd transition. For these reasons, an analysis of the $^1S_0 \to {}^3D_2$ transition is not included here. 

\section{Interrogation scheme}
Alkaline earth atoms are loaded from an oven or a getter \cite{Bridge2009} into a compact ultra-high-vacuum chamber pumped by an ion pump. A standard MOT configuration is used for laser cooling and trapping on the strong $4s^2 \ ^1S_0 \to 4s 4p \ {}^1P_1$ transition ($\gamma_1/2\pi \approx$ 34 MHz) (using a laser L1, 423 nm). There is a very small rate of shelving into the $^1D_2$ state ($10^{-5}$/cycle), out of which atoms can be repumped back into the cooling cycle using a laser (L2, 672 nm) tuned to the strongly allowed $4s 3d \ ^1D_2 \to 4s 5p {}^1P_1$ transition ($\gamma_2/2\pi \approx$ 2 MHz). After loading the trap, the cooling lasers and MOT magnetic fields are switched off to avoid light shifts, optical pumping and Zeeman shifts during the interrogation phase. Two counter-propagating beams, derived from the same narrow linewidth laser oscillator (LO) (L3, 916 nm), are then used to drive the \sd clock transition ($\gamma_3/2\pi \approx$ 40 Hz) in a Rabi or Ramsey scheme. We assume that a sufficiently long interrogation sequence is used ($\geq$6.3 ms for a Ramsey sequence) so that the measured linewidth is limited by the natural linewidth of the transition.

The excitation probability is measured using laser-induced cycling fluorescence on the $^1D_2 \leftrightarrow 4s 4f \ {}^1F_3$ cycling transition (L4, 488 nm) \footnote{D. DeMille, private communication (2014)}, or by measuring the drop in the MOT's fluorescence when the laser L1 (but not L2) is switched back on. Collection of fluorescence on these cycling transitions means that small-solid-angle detectors can be conveniently used while still obtaining near-unity detection efficiency, at a noise level limited by quantum projection noise in the detection process. During a 10 ms-long interrogation + detection sequence, the atoms move $\sim$ 7 mm. Therefore a large fraction of them can be recaptured by the MOT and re-used for the next cycle.

Assuming that $N=10^6$ atoms can be interrogated in the MOT and detected with shot-noise-limited sensitivity, the frequency resolution of the clock, with a natural lifetime $T$ and an integration time $\tau$, is $\delta \nu = 1/2\pi \sqrt{N T \tau}$. Using $T =$ 2 ms, this evaluates to a fractional frequency resolution $\delta \nu/\nu = 10^{-17}/\sqrt{\tau(s)}$. Even allowing that this might be reduced due to the duty cycle of the interrogation, this is an extremely attractive sensitivity for a secondary standard, which must be capable of quick comparisons with primary standards and other frequency references. Further, the fast cycle time means that Dick effect noise is less important -- the flywheel for the laser's frequency only needs to carry it over for $\sim$ 10 ms, until the next interrogation -- leading to relaxed requirements on the LO. Calculations using realistic parameters for a MOT indicate that a two-photon Rabi frequency $\Omega_\mrm{eff}$ = $2\pi \times$ 300 Hz can be achieved with 1 W of LO power, commensurate with a modest power build-up cavity around the MOT that is fed by a 916 nm diode laser. All the lasers can be derived from laser diodes, and the scheme is compatible with a low-mass, low-power apparatus. We also note that this scheme lends itself quite naturally to methods that attempt to push beyond the standard quantum limit, using atom-cavity interactions \cite{Schleier-Smith2010} 	and/or non-classical light sources.

\begin{figure}{
\centering
\includegraphics[width=\columnwidth]{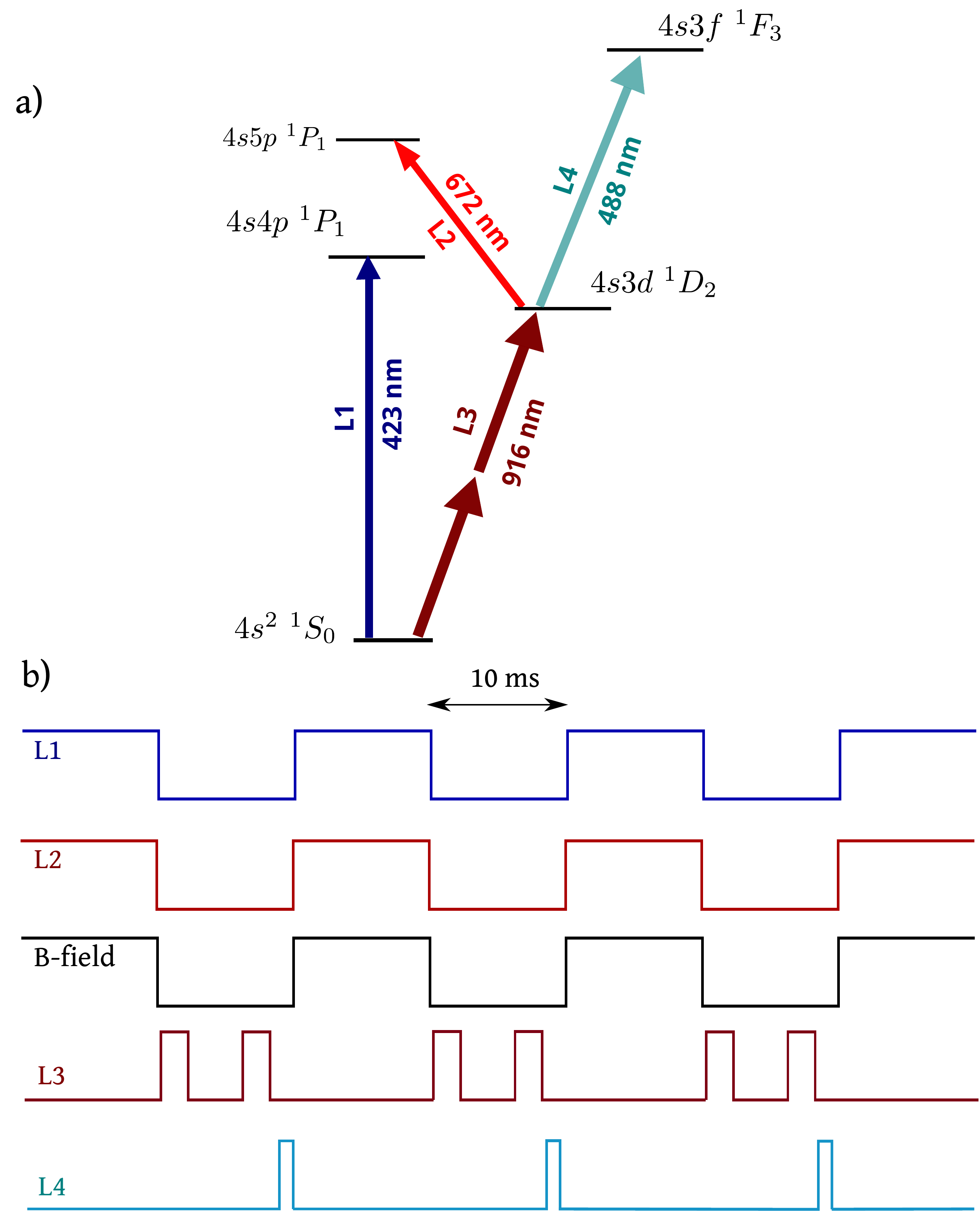}
\caption{a) Energy levels of neutral calcium, showing the atomic transitions involved in the magneto-optical trapping and interrogation scheme. b) The timing sequence of the MOT lasers (L1 \& L2), B-field, clock laser (L3) and detection laser (L4). Exciting the clock transition with two counter-propagating L3 photons leads to Doppler- and recoil-free excitation, which is probed with fluorescence induced by L4.}}
\label{fig:energyLevels}
\end{figure}

\section{Systematic effects}
\subsection{2nd order Doppler shift}
Using $\sim$ 2 mK for the temperature of a single stage 423 nm MOT, the root-mean-square velocity of the calcium atoms is $v_\mrm{rms} \approx$ 70 cm/s. This leads to a 2nd-order Doppler shift whose fractional size is 
\begin{equation}
\frac{\delta \nu}{\nu} = -\frac{1}{2} \frac{v_\mrm{rms}^2}{c^2} = -2.7 \times 10^{-18}.
\end{equation}
This rather small number implies that the stability of the temperature of the atoms in the MOT will not affect the operation of the frequency standard to any relevant degree.

\subsection{Collisions}
Assuming a MOT density $n_\mrm{MOT} = 10^9$/cm$^3$ and a collision cross section $\sigma \simeq 10^{-14}$ cm$^2$, the estimated mean free time between collisions is $\tau_\mrm{coll} \simeq$ 1500 s. This is significantly larger than the expected cycle time of the interrogation. Conservatively assuming $\sim \pi$ rad phase shift per collision, the (fractional) collisional frequency shift evaluates to $\sim 3 \times 10^{-18}$. Therefore we consider it likely that collisional effects can be controlled or calibrated at the level of $\leq 10^{-16}$.

\subsection{Electric shifts}
The DC electric and light shifts were evaluated numerically, using the known transition rates (and dipole matrix elements derived from them) for the singlet states up to $4s 5p$ \cite{Hansen1999}. The calculated DC polarizability  of the $4s^2 \ ^1S_0$ state is 75, $a_0^3$ and that of the $4s 3d \ ^1D_2$ state is 32 $a_0^3$. The resulting DC electric shift of the transition is $\delta \nu_\mrm{E} \approx -\Delta \alpha_\mrm{DC} \Esca^2 \approx$ +13 mHz/(V/cm)$^2$, $(\delta \nu/\nu)_\mrm{E} \approx 2 \times 10^{-17}$/(V/cm)$^2$.

\begin{figure}{
\centering
\includegraphics[width=\columnwidth]{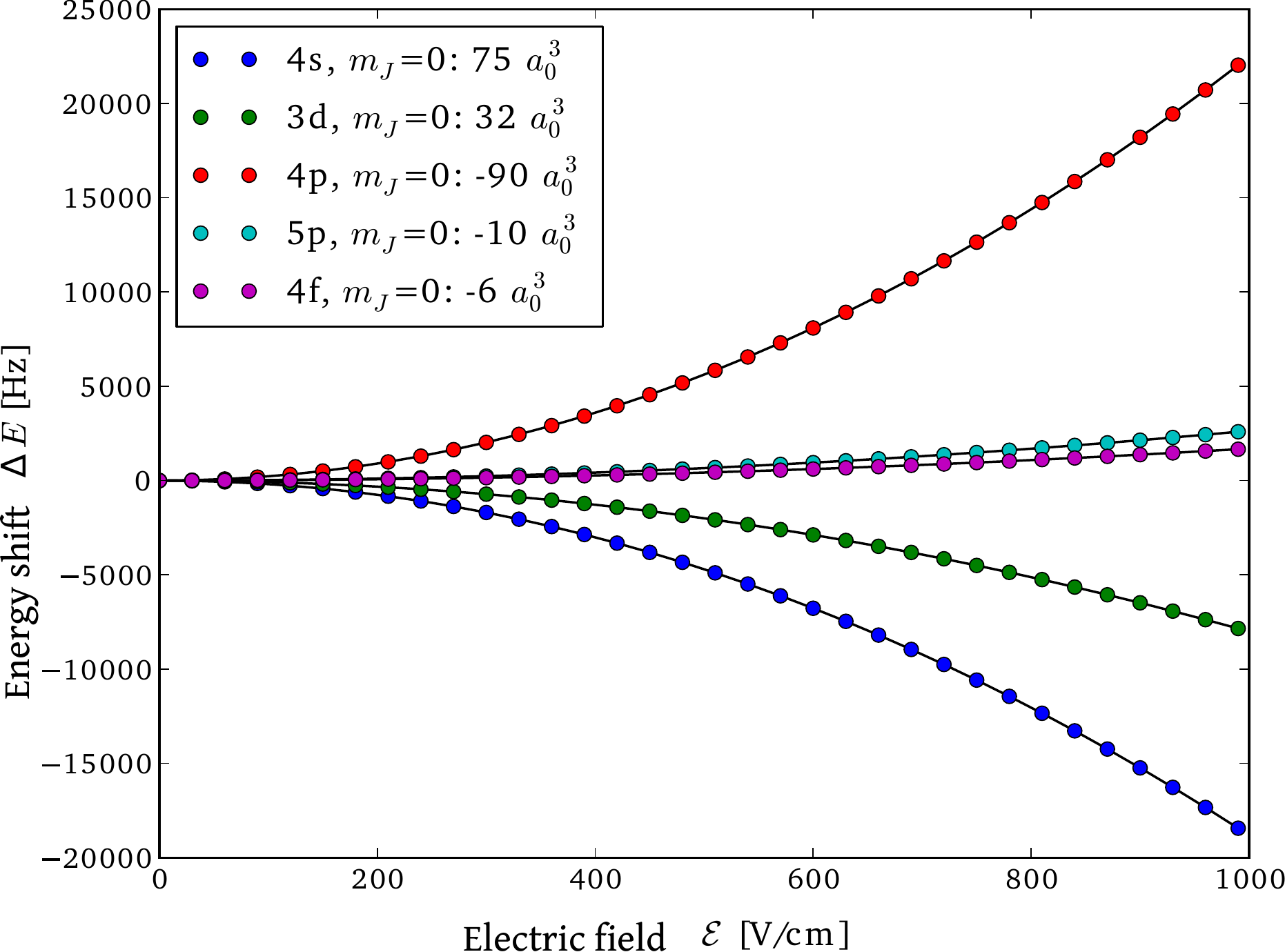}
\caption{Calculated DC electric-field-induced energy shifts of the singlet $m=0$ levels in calcium. The calculated energy shifts are are fit to a quadratic curve to obtain the DC polarizability $\alpha_\mrm{DC}$.}}
\label{fig:electric-shifts}
\end{figure}

The light shift of the clock transition due to photons at 916 nm was evaluated using dressed states (using both the co- and counter-rotating components). It is equal to $\Delta E_\mrm{LS} \simeq 12$ Hz/(W/cm$^2$) for $\hat{z}$-polarized light. 

Note that for many of the applications of a secondary standard, it is sufficient that the frequency (shifts) be stable and repeatable. However, there are also ways to improve the absolute accuracy of the frequency standard by cancelling the light shift: a) the light shift and the (spatial) excitation profile scale in the same way with the laser intensity. Once the light shift is calibrated, it can be applied as a correction that is proportional to the (measured) excitation probability. cf.\ \cite{Huntemann2012a} for an example of a highly forbidden transition, where the light shift is cancelled by extrapolating the laser power. b) There are variants of the Ramsey pulse sequence that have been applied to forbidden clock transitions (``hyper-Ramsey'' pulse sequences) \cite{Yudin2010}, where the light shift can be cancelled at the expense of some complexity in the interrogation pulse sequence.

\subsection{Black-body radiation shift}
The BBR shift can be approximately estimated from the DC electric polarizabilities, since the relevant atomic transitions are well to the blue of the thermal photon distribution:
\begin{equation}
\delta \nu_\mrm{BBR} \approx -\frac{2}{15}(\alpha \pi)^3 T^4 \ [\alpha_\mrm{DC}({}^1D_2) - \alpha_\mrm{s}({}^1S_0)].
\end{equation}
(Here $\alpha$ is the fine structure constant, $T$ is the temperature in atomic units, and the value for $\delta \nu_\mrm{BBR}$ is also in atomic units.) Using the above-calculated DC polarizabilities, this yields $(\delta \nu/\nu)_\mrm{BBR} \approx 0.6 \times 10^{-15}$. This BBR shift only needs to be evaluated to $\sim$ 10\% accuracy, to obtain $\leq 10^{-16}$ fractional accuracy of the frequency standard at room temperature. 

\subsection{Magnetic shifts}
The clock states are both $m=0$ states, and the most abundant calcium isotope has zero nuclear magnetic moment. The dominant source of magnetic shifts of the clock transition is a 2nd-order Zeeman shift due to the $\Bsca$-field-induced mixing of the $^1D_2$ and $^3D_{1,3}$ states. Assuming a $\sim$1 $\mu_B$ matrix element for this spin flip transition, and using the energy difference $\Delta \approx$ 50 THz between these states, the magnetic shift coefficient for the clock transition is estimated to be  $\delta \nu_\mrm{B} \approx$ 50 mHz/G$^2$. The fractional shift is $\Big(\frac{\delta \nu}{\nu}\Big)_B \approx 2 \times 10^{-16}$/G$^2$. This bodes well for achieving a fractional frequency accuracy better than $10^{-16}$ using field cancellation coils and/or simple magnetic shielding.

\begin{table}
\centering
\begin{tabular}{lll}
\hline 
Systematic 	& Parameter	&  Fractional frequency \\
effect & control range  & shift, $\frac{\delta \nu}{\nu}$ ($10^{-17}$)\\
\hline
2$^\mrm{nd}$ order Doppler \ \ \ \ \ \ & $v_\mrm{rms} \lesssim$ 1 m/s & 0.5 \\ 
Electric shift & $\Esca \lesssim$ 0.1 V/cm \ \ \ \ \ \ \ & 0.02 \\ 
Light shift & $\frac{\delta P}{P} \lesssim 10^{-3}$ & 1.8\\ 
BBR shift & $\frac{\delta \alpha_\mrm{s}}{\alpha_\mrm{s}} \lesssim$ 5\% & 3  \\ 
Magnetic shift & $\Bsca \lesssim$ 0.1 G & 0.2 \\ 
\hline 
\end{tabular} 
\caption{\small \emph{Estimated contributions to the systematic shifts in a calcium-MOT-based optical frequency reference, using realistic ranges for parameters that can be controlled or calculated.}}
\end{table}

\section{Summary}
A scheme to construct an optical frequency standard has been described, based on the \sd two-photon transition in calcium atoms trapped in a magneto-optical trap. Using a simple apparatus, it is capable of achieving statistical sensitivity and systematic immunity at the level of $\leq$ 1 part in  $10^{16}$. The dominant systematic effects that are likely to affect the operation of the frequency standard are listed in Table I. An implementation of this scheme in a compact calcium MOT could lead to robust and portable frequency standards, with a multitude of potential applications in time \& frequency transfer, geophysics and precision measurements. We have begun the construction of a prototype device.

\section*{Acknowledgments}
I am grateful to Eric Hessels and Dave DeMille for their encouragement and helpful suggestions. I have benefited greatly from conversations with Stephan Falke and Uwe Sterr. I thank Eric Hudson, Wes Campbell and the ACME collaboration for the loan of equipment during a preliminary experiment. This work is supported by a Society in Science Branco Weiss Fellowship, administered by the ETH Zurich.

\bibliography{calcium_clock}

\begin{thebibliography}{16}%
\makeatletter
\providecommand \@ifxundefined [1]{%
 \@ifx{#1\undefined}
}%
\providecommand \@ifnum [1]{%
 \ifnum #1\expandafter \@firstoftwo
 \else \expandafter \@secondoftwo
 \fi
}%
\providecommand \@ifx [1]{%
 \ifx #1\expandafter \@firstoftwo
 \else \expandafter \@secondoftwo
 \fi
}%
\providecommand \natexlab [1]{#1}%
\providecommand \enquote  [1]{``#1''}%
\providecommand \bibnamefont  [1]{#1}%
\providecommand \bibfnamefont [1]{#1}%
\providecommand \citenamefont [1]{#1}%
\providecommand \href@noop [0]{\@secondoftwo}%
\providecommand \href [0]{\begingroup \@sanitize@url \@href}%
\providecommand \@href[1]{\@@startlink{#1}\@@href}%
\providecommand \@@href[1]{\endgroup#1\@@endlink}%
\providecommand \@sanitize@url [0]{\catcode `\\12\catcode `\$12\catcode
  `\&12\catcode `\#12\catcode `\^12\catcode `\_12\catcode `\%12\relax}%
\providecommand \@@startlink[1]{}%
\providecommand \@@endlink[0]{}%
\providecommand \url  [0]{\begingroup\@sanitize@url \@url }%
\providecommand \@url [1]{\endgroup\@href {#1}{\urlprefix }}%
\providecommand \urlprefix  [0]{URL }%
\providecommand \Eprint [0]{\href }%
\providecommand \doibase [0]{http://dx.doi.org/}%
\providecommand \selectlanguage [0]{\@gobble}%
\providecommand \bibinfo  [0]{\@secondoftwo}%
\providecommand \bibfield  [0]{\@secondoftwo}%
\providecommand \translation [1]{[#1]}%
\providecommand \BibitemOpen [0]{}%
\providecommand \bibitemStop [0]{}%
\providecommand \bibitemNoStop [0]{.\EOS\space}%
\providecommand \EOS [0]{\spacefactor3000\relax}%
\providecommand \BibitemShut  [1]{\csname bibitem#1\endcsname}%
\let\auto@bib@innerbib\@empty
\bibitem [{\citenamefont {Gill}(2005)}]{Gill2005}%
  \BibitemOpen
  \bibfield  {author} {\bibinfo {author} {\bibfnamefont {P.}~\bibnamefont
  {Gill}},\ }\href {\doibase 10.1088/0026-1394/42/3/S13} {\bibfield  {journal}
  {\bibinfo  {journal} {Metrologia}\ }\textbf {\bibinfo {volume} {42}},\
  \bibinfo {pages} {S125} (\bibinfo {year} {2005})}\BibitemShut {NoStop}%
\bibitem [{\citenamefont {Poli}\ \emph {et~al.}(2014)\citenamefont {Poli},
  \citenamefont {Oates}, \citenamefont {Gill},\ and\ \citenamefont
  {Tino}}]{Poli2014}%
  \BibitemOpen
  \bibfield  {author} {\bibinfo {author} {\bibfnamefont {N.}~\bibnamefont
  {Poli}}, \bibinfo {author} {\bibfnamefont {C.}~\bibnamefont {Oates}},
  \bibinfo {author} {\bibfnamefont {P.}~\bibnamefont {Gill}}, \ and\ \bibinfo
  {author} {\bibfnamefont {G.}~\bibnamefont {Tino}},\ }\href
  {http://arxiv.org/abs/1401.2378} {\bibfield  {journal} {\bibinfo  {journal}
  {arXiv:1401.2378}\ }\textbf {\bibinfo {volume} {36}},\ \bibinfo {pages} {555}
  (\bibinfo {year} {2014})},\ \Eprint {http://arxiv.org/abs/arXiv:1401.2378v1}
  {arXiv:arXiv:1401.2378v1} \BibitemShut {NoStop}%
\bibitem [{\citenamefont {Ludlow}\ \emph {et~al.}(2014)\citenamefont {Ludlow},
  \citenamefont {Boyd}, \citenamefont {Ye}, \citenamefont {Peik},\ and\
  \citenamefont {Schmidt}}]{Ludlow2014}%
  \BibitemOpen
  \bibfield  {author} {\bibinfo {author} {\bibfnamefont {A.}~\bibnamefont
  {Ludlow}}, \bibinfo {author} {\bibfnamefont {M.}~\bibnamefont {Boyd}},
  \bibinfo {author} {\bibfnamefont {J.}~\bibnamefont {Ye}}, \bibinfo {author}
  {\bibfnamefont {E.}~\bibnamefont {Peik}}, \ and\ \bibinfo {author}
  {\bibfnamefont {P.}~\bibnamefont {Schmidt}},\ }\href@noop {} {\bibfield
  {journal} {\bibinfo  {journal} {arXiv}\ }\textbf {\bibinfo {volume}
  {arXiv:1407.3493}} (\bibinfo {year} {2014})}\BibitemShut {NoStop}%
\bibitem [{\citenamefont {Margolis}(2014)}]{Margolis2014}%
  \BibitemOpen
  \bibfield  {author} {\bibinfo {author} {\bibfnamefont {H.}~\bibnamefont
  {Margolis}},\ }\href {\doibase 10.1038/nphys2834} {\bibfield  {journal}
  {\bibinfo  {journal} {Nat. Phys.}\ }\textbf {\bibinfo {volume} {10}},\
  \bibinfo {pages} {82} (\bibinfo {year} {2014})}\BibitemShut {NoStop}%
\bibitem [{\citenamefont {Parker}(2012)}]{Parker2012}%
  \BibitemOpen
  \bibfield  {author} {\bibinfo {author} {\bibfnamefont {T.~E.}\ \bibnamefont
  {Parker}},\ }\href {\doibase 10.1063/1.3682002} {\bibfield  {journal}
  {\bibinfo  {journal} {Rev. Sci. Instrum.}\ }\textbf {\bibinfo {volume}
  {83}},\ \bibinfo {pages} {021102} (\bibinfo {year} {2012})}\BibitemShut
  {NoStop}%
\bibitem [{\citenamefont {McFerran}\ \emph {et~al.}(2005)\citenamefont
  {McFerran}, \citenamefont {Ivanov},\ and\ \citenamefont
  {Bartels}}]{McFerran2005}%
  \BibitemOpen
  \bibfield  {author} {\bibinfo {author} {\bibfnamefont {J.}~\bibnamefont
  {McFerran}}, \bibinfo {author} {\bibfnamefont {E.}~\bibnamefont {Ivanov}}, \
  and\ \bibinfo {author} {\bibfnamefont {A.}~\bibnamefont {Bartels}},\ }\href
  {\doibase 10.1049/el} {\bibfield  {journal} {\bibinfo  {journal} {Electron.
  Lett.}\ }\textbf {\bibinfo {volume} {41}},\ \bibinfo {pages} {40} (\bibinfo
  {year} {2005})}\BibitemShut {NoStop}%
\bibitem [{\citenamefont {Chou}\ \emph {et~al.}(2010)\citenamefont {Chou},
  \citenamefont {Hume}, \citenamefont {Rosenband},\ and\ \citenamefont
  {Wineland}}]{Chou2010}%
  \BibitemOpen
  \bibfield  {author} {\bibinfo {author} {\bibfnamefont {C.~W.}\ \bibnamefont
  {Chou}}, \bibinfo {author} {\bibfnamefont {D.~B.}\ \bibnamefont {Hume}},
  \bibinfo {author} {\bibfnamefont {T.}~\bibnamefont {Rosenband}}, \ and\
  \bibinfo {author} {\bibfnamefont {D.~J.}\ \bibnamefont {Wineland}},\ }\href
  {\doibase 10.1126/science.1192720} {\bibfield  {journal} {\bibinfo  {journal}
  {Science (80-. ).}\ }\textbf {\bibinfo {volume} {329}},\ \bibinfo {pages}
  {1630} (\bibinfo {year} {2010})}\BibitemShut {NoStop}%
\bibitem [{\citenamefont {Smarr}\ \emph {et~al.}(1983)\citenamefont {Smarr},
  \citenamefont {Vessot}, \citenamefont {Lundquist}, \citenamefont {Decher},\
  and\ \citenamefont {Piran}}]{Smarr1983}%
  \BibitemOpen
  \bibfield  {author} {\bibinfo {author} {\bibfnamefont {L.~L.}\ \bibnamefont
  {Smarr}}, \bibinfo {author} {\bibfnamefont {R.~F.~C.}\ \bibnamefont
  {Vessot}}, \bibinfo {author} {\bibfnamefont {C.~A.}\ \bibnamefont
  {Lundquist}}, \bibinfo {author} {\bibfnamefont {R.}~\bibnamefont {Decher}}, \
  and\ \bibinfo {author} {\bibfnamefont {T.}~\bibnamefont {Piran}},\ }\href
  {\doibase 10.1007/BF00762473} {\bibfield  {journal} {\bibinfo  {journal}
  {Gen. Relativ. Gravit.}\ }\textbf {\bibinfo {volume} {15}},\ \bibinfo {pages}
  {129} (\bibinfo {year} {1983})}\BibitemShut {NoStop}%
\bibitem [{\citenamefont {Armstrong}(2006)}]{Armstrong2006}%
  \BibitemOpen
  \bibfield  {author} {\bibinfo {author} {\bibfnamefont {J.}~\bibnamefont
  {Armstrong}},\ }\href
  {http://www.emis.ams.org/journals/LRG/Articles/lrr-2006-1/?affiliation}
  {\bibfield  {journal} {\bibinfo  {journal} {Living Rev. Relativ.}\ }
  (\bibinfo {year} {2006})}\BibitemShut {NoStop}%
\bibitem [{\citenamefont {Hall}\ \emph {et~al.}(1989)\citenamefont {Hall},
  \citenamefont {Zhu},\ and\ \citenamefont {Buch}}]{Hall1989}%
  \BibitemOpen
  \bibfield  {author} {\bibinfo {author} {\bibfnamefont {J.}~\bibnamefont
  {Hall}}, \bibinfo {author} {\bibfnamefont {M.}~\bibnamefont {Zhu}}, \ and\
  \bibinfo {author} {\bibfnamefont {P.}~\bibnamefont {Buch}},\ }\href
  {http://www.opticsinfobase.org/abstract.cfm?uri=josab-6-11-2194} {\bibfield
  {journal} {\bibinfo  {journal} {Josa B}\ }\textbf {\bibinfo {volume} {6}},\
  \bibinfo {pages} {2194} (\bibinfo {year} {1989})}\BibitemShut {NoStop}%
\bibitem [{\citenamefont {Bridge}\ and\ \citenamefont
  {Millen}(2009)}]{Bridge2009}%
  \BibitemOpen
  \bibfield  {author} {\bibinfo {author} {\bibfnamefont {E.}~\bibnamefont
  {Bridge}}\ and\ \bibinfo {author} {\bibfnamefont {J.}~\bibnamefont
  {Millen}},\ }\href {\doibase 10.1063/1.3036980} {\bibfield  {journal}
  {\bibinfo  {journal} {Rev. Sci. Inst.}\ }\textbf {\bibinfo {volume} {44}},\
  \bibinfo {pages} {0} (\bibinfo {year} {2009})}\BibitemShut {NoStop}%
\bibitem [{Note1()}]{Note1}%
  \BibitemOpen
  \bibinfo {note} {D. DeMille, private communication (2014)}\BibitemShut
  {NoStop}%
\bibitem [{\citenamefont {Schleier-Smith}\ \emph {et~al.}(2010)\citenamefont
  {Schleier-Smith}, \citenamefont {Leroux},\ and\ \citenamefont
  {Vuleti\ifmmode~\acute{c}\else \'{c}\fi{}}}]{Schleier-Smith2010}%
  \BibitemOpen
  \bibfield  {author} {\bibinfo {author} {\bibfnamefont {M.~H.}\ \bibnamefont
  {Schleier-Smith}}, \bibinfo {author} {\bibfnamefont {I.~D.}\ \bibnamefont
  {Leroux}}, \ and\ \bibinfo {author} {\bibfnamefont {V.}~\bibnamefont
  {Vuleti\ifmmode~\acute{c}\else \'{c}\fi{}}},\ }\href {\doibase
  10.1103/PhysRevLett.104.073604} {\bibfield  {journal} {\bibinfo  {journal}
  {Phys. Rev. Lett.}\ }\textbf {\bibinfo {volume} {104}},\ \bibinfo {pages}
  {073604} (\bibinfo {year} {2010})}\BibitemShut {NoStop}%
\bibitem [{\citenamefont {Hansen}\ and\ \citenamefont
  {Laughlin}(1999)}]{Hansen1999}%
  \BibitemOpen
  \bibfield  {author} {\bibinfo {author} {\bibfnamefont {J.}~\bibnamefont
  {Hansen}}\ and\ \bibinfo {author} {\bibfnamefont {C.}~\bibnamefont
  {Laughlin}},\ }\href {http://iopscience.iop.org/0953-4075/32/9/305}
  {\bibfield  {journal} {\bibinfo  {journal} {J. Phys. B At. Mol. Phys.}\
  }\textbf {\bibinfo {volume} {2099}} (\bibinfo {year} {1999})}\BibitemShut
  {NoStop}%
\bibitem [{\citenamefont {Huntemann}\ \emph {et~al.}(2012)\citenamefont
  {Huntemann}, \citenamefont {Okhapkin}, \citenamefont {Lipphardt},
  \citenamefont {Weyers}, \citenamefont {Tamm},\ and\ \citenamefont
  {Peik}}]{Huntemann2012a}%
  \BibitemOpen
  \bibfield  {author} {\bibinfo {author} {\bibfnamefont {N.}~\bibnamefont
  {Huntemann}}, \bibinfo {author} {\bibfnamefont {M.}~\bibnamefont {Okhapkin}},
  \bibinfo {author} {\bibfnamefont {B.}~\bibnamefont {Lipphardt}}, \bibinfo
  {author} {\bibfnamefont {S.}~\bibnamefont {Weyers}}, \bibinfo {author}
  {\bibfnamefont {C.}~\bibnamefont {Tamm}}, \ and\ \bibinfo {author}
  {\bibfnamefont {E.}~\bibnamefont {Peik}},\ }\href {\doibase
  10.1103/PhysRevLett.108.090801} {\bibfield  {journal} {\bibinfo  {journal}
  {Phys. Rev. Lett.}\ }\textbf {\bibinfo {volume} {108}},\ \bibinfo {pages}
  {090801} (\bibinfo {year} {2012})}\BibitemShut {NoStop}%
\bibitem [{\citenamefont {Yudin}\ \emph {et~al.}(2010)\citenamefont {Yudin},
  \citenamefont {Taichenachev}, \citenamefont {Oates}, \citenamefont {Barber},
  \citenamefont {Lemke}, \citenamefont {Ludlow}, \citenamefont {Sterr},
  \citenamefont {Lisdat},\ and\ \citenamefont {Riehle}}]{Yudin2010}%
  \BibitemOpen
  \bibfield  {author} {\bibinfo {author} {\bibfnamefont {V.~I.}\ \bibnamefont
  {Yudin}}, \bibinfo {author} {\bibfnamefont {a.~V.}\ \bibnamefont
  {Taichenachev}}, \bibinfo {author} {\bibfnamefont {C.~W.}\ \bibnamefont
  {Oates}}, \bibinfo {author} {\bibfnamefont {Z.~W.}\ \bibnamefont {Barber}},
  \bibinfo {author} {\bibfnamefont {N.~D.}\ \bibnamefont {Lemke}}, \bibinfo
  {author} {\bibfnamefont {a.~D.}\ \bibnamefont {Ludlow}}, \bibinfo {author}
  {\bibfnamefont {U.}~\bibnamefont {Sterr}}, \bibinfo {author} {\bibfnamefont
  {C.}~\bibnamefont {Lisdat}}, \ and\ \bibinfo {author} {\bibfnamefont
  {F.}~\bibnamefont {Riehle}},\ }\href {\doibase 10.1103/PhysRevA.82.011804}
  {\bibfield  {journal} {\bibinfo  {journal} {Phys. Rev. A}\ }\textbf {\bibinfo
  {volume} {82}},\ \bibinfo {pages} {011804} (\bibinfo {year}
  {2010})}\BibitemShut {NoStop}%
\end{thebibliography}%
\end{document}